\documentclass[PAPER, cernpreprint, UKenglish, texmf, orcidlogo]{atlasdoc}

\usepackage{atlaspackage}
\usepackage{atlasbiblatex}
 
\usepackage{atlasphysics}

\usepackage{multirow}
 
\graphicspath{{logos/}{figures/}}
 
\usepackage{ANA-HIGG-2020-03-PAPER-defs}
 
\AtlasTitle{Probing the $CP$ nature of the top--Higgs Yukawa coupling in $t\bar{t}H$ and $tH$ events with $H \rightarrow b\bar{b}$ decays using the ATLAS detector at the LHC}
 
\PreprintIdNumber{CERN-EP-2022-208}
\AtlasAbstract{
The $CP$ properties of the coupling between the Higgs boson and the top quark are investigated using 139~fb$^{-1}$\ of proton--proton collision data
recorded by the ATLAS experiment at the LHC at a centre-of-mass energy of $\sqrt{s}=13$~TeV.
The $CP$ structure of the top quark--Higgs boson Yukawa coupling is probed
in events with a Higgs boson decaying into a pair of $b$-quarks and produced in association with either a pair of top quarks, $t\bar{t}H$, or a single top quark, $tH$.
Events containing one or two electrons or muons are used for the measurement. 
Multivariate techniques are used to select regions enriched in $t\bar{t}H$ and $tH$ events, where dedicated $CP$-sensitive observables are exploited.
In an extension of the Standard Model (SM) with a $CP$-odd admixture in the top--Higgs Yukawa coupling,
the mixing angle between $CP$-even and $CP$-odd couplings is measured to be $\alpha=11\,^{\circ+52^{\circ}}_{~-73^{\circ}}$,
compatible with the SM prediction corresponding to $\alpha=0$.
 
}
 
\author{The ATLAS Collaboration}
 
\AtlasRefCode{HIGG-2020-03-001}
 
\AtlasJournalRef{\href{https://www.sciencedirect.com/science/article/pii/S0370269324000285/pdf}{Phys.\ Lett.\ B.\ 849 (2024) 138469}}
\AtlasDOI{10.1016/j.physletb.2024.138469}
 
\AtlasDate{\today}
 
\AtlasHeading{ATLAS Paper Draft}

\hypersetup{pdftitle={ATLAS document},pdfauthor={The ATLAS Collaboration}}
 
\begin{document}
 
\maketitle

\section{Introduction}

 
Since the observation of the Higgs boson at the LHC~\cite{HIGG-2012-27,CMS-HIG-12-036},
its properties have been studied in great detail.
In particular, the observation of the Higgs boson production in association with a top-quark pair, \ttH~\cite{HIGG-2018-13,CMS-HIG-17-035},
provides direct experimental access to the top-quark Yukawa coupling at the tree-level.
The increasing datasets at the LHC have recently allowed the ATLAS and CMS Collaborations
to probe the charge-conjugation and parity (\(CP\)) properties of this coupling using
\ttH~events in different decay channels~\cite{HIGG-2019-01,CMS-HIG-19-013,CMS-HIG-21-006}.
This letter reports on a study
of the \(CP\) properties of the top-quark Yukawa coupling using \ttH\ and \tH\ production, in the $H\rightarrow b\bar{b}$ decay channel.
The analysis targets final states where at least one top quark decays semi-leptonically to electrons or muons.
It uses $\sqrt{s}=13$~\TeV\ $pp$ collision data recorded by the ATLAS experiment during Run~2,
corresponding to an integrated luminosity of 139~\ifb.

The Standard Model (SM) predicts the Higgs boson to be a scalar particle with quantum numbers $J^{CP} = 0^{++}$. Considering the possibility of beyond the Standard Model (BSM) couplings, a \(CP\)-odd component of the vector-boson couplings to the Higgs boson is naturally suppressed by the scale at which new physics would become relevant~\cite{BUCHMULLER1986621,Hankele_2006}.
This suppression does not happen for Yukawa couplings, where \(CP\)-odd Higgs--fermion couplings may be significant already at tree level~\cite{Fontes:2015}.
One of the first ATLAS and CMS measurements have excluded the pure $J^{P} = 0^{-}$ hypothesis by more than 95\% CL
using $H\rightarrow \gamma \gamma$, $H\rightarrow ZZ^*$ and $H\rightarrow WW^*$ decays~\cite{HIGG-2013-01,CMS-HIG-12-041}.
Dedicated searches for $CP$-mixed couplings between the Higgs boson and vector bosons set stringent limits on the $CP$-odd components~\cite{HIGG-2013-06,HIGG-2018-14,HIGG-2016-21,HIGG-2016-22,CMS-HIG-14-035,CMS-HIG-17-011,CMS-HIG-18-002,CMS-HIG-20-007,HIGG-2018-30}.
Analyses of \ttH~events with $H\rightarrow \gamma \gamma$ decays~\cite{HIGG-2019-01,CMS-HIG-19-013} and in the multilepton final state~\cite{CMS-HIG-21-006} have also excluded pure \(CP\)-odd top--Higgs couplings at more than a $3~\sigma$ significance. But mixing of \(CP\)-odd and \(CP\)-even states has not been ruled out and is worth investigating.
The observation of a non-zero \(CP\)-odd coupling component would in fact signal the existence of BSM physics, and open up the possibility of \(CP\)-violation in the Higgs sector~\cite{Gunion:1996xu,Ellis:2014hw,He:2015gu,Boudjema:2015nda}.
Such a new source of \(CP\) violation could play a fundamental role in explaining the matter--antimatter asymmetry of the universe.
This analysis targets \ttH\ and \tH\ events, which are sensitive to the top--Higgs coupling including any potential $CP$-mixing at the tree-level.
This avoids the need for assumptions about the influence of BSM effects which may be present in other, more indirect measurements~\cite{Demartin:2014fia,Bahl:2020, Goncalves:2021}. In particular, current limits on electron and neutron electrical dipole moments place indirect model-dependent constraints on a possible pseudoscalar component of the top-quark Yukawa coupling~\cite{Brod:2013edm,acme:2018edm,nEDM2020edm}.

The top--Higgs interaction can be extended beyond the SM as~\cite{Demartin:2014fia}:
\begin{equation}{\cal L}_{\ttH}=-\kprime y_t\phi\bar{\psi_t}(\cos{\alpha} + i\gamma_5\sin{\alpha})\psi_t\,,
\label{eq:lagrangian}
\end{equation}
where $y_t$ is the SM Yukawa coupling strength, modified by a coupling modifier $\kprime$; $\alpha$ is the \(CP\)-mixing angle;
$\phi$ is the Higgs field; $\psi_t$ and $\bar{\psi_t}$ are top-quark spinor fields and $\gamma_5$ is a Dirac matrix.
The term containing $\gamma_5$ corresponds to a pseudoscalar component.
The above expression reduces to the SM case for $\kprime = 1$ and $\alpha=0$.
An anomalous value of $\alpha$ would produce an admixture with a pseudoscalar coupling ($J^{CP}=0^{+-}$) and change  the differential cross-section relative to the SM expectation, while a variation of $\kprime$ would induce a change in the total cross-section~\cite{Gunion:1996xu,Buckley:2015vsa,Goncalves:2018agy,Azevedo:2017qiz,Ferroglia:2019qjy}.

This study measures simultaneously the values of \kprime~and $\alpha$ with a binned profile likelihood fit to data, exploiting dedicated $CP$-sensitive observables.
It closely follows  a recent analysis optimised for the measurement of the $t\bar{t}H(\to bb)$\ production cross-section~\cite{HIGG-2020-23}.
This analysis studies an identical phase space using the same physics object definitions and a similar methodology for event selection and evaluation of systematic uncertainties. 
A notable exception is that this analysis considers both the \ttH\ and \tH\ production modes as signals. No attempt was made to optimise the analysis strategy for the \tH\ signal, as its small yield makes this channel relevant only in one analysis region.
Other noteworthy differences with respect to the analysis documented in Ref.~\cite{HIGG-2020-23} are detailed in the text.
These include the definition of signal regions, the signal-background discrimination strategy and a few details in the definition of systematic uncertainties in signal and background modelling.
In the case of \tH\ production, the destructive interference between the diagrams with $t$--$H$ and $W$--$H$ couplings leads to the negligible  \tH\ production cross-section in the SM.
Any change in the relative $t$--$H$ and $W$--$H$ coupling strengths would result in a rapid increase in the cross-section.
Considering the Lagrangian density in Eqn.~\ref{eq:lagrangian}, the \tH\ production cross-section is expected to grow for values of the mixing angle $\alpha$ different from zero~\cite{Ellis:2014hw}.
An opposite and less pronounced dependence exists for the \ttH\ cross-section.
The ratio of \tH\ to \ttH\ cross-sections varies from 0.06 in the SM scenario to more than 1.2 in the pure \(CP\)-odd scenario~\cite{Ellis:2014hw}.
For the present measurement, the $H\rightarrow b\bar{b}$ branching ratio is assumed to be equal to its SM value
of 58.2\% $\pm$ 0.5\%~\cite{deFlorian:2016spz}.


\section{The ATLAS experiment}

 
\newcommand{\AtlasCoordFootnote}{
ATLAS uses a right-handed coordinate system with its origin at the nominal interaction point (IP)
in the centre of the detector and the \(z\)-axis along the beam pipe.
The \(x\)-axis points from the IP to the centre of the LHC ring,
and the \(y\)-axis points upwards.
Cylindrical coordinates \((r,\phi)\) are used in the transverse plane,
\(\phi\) being the azimuthal angle around the \(z\)-axis.
The momentum component in the transverse plane is referred to as the transverse momentum (\pt).
The pseudorapidity is defined in terms of the polar angle \(\theta\) as \(\eta = -\ln \tan(\theta/2)\).
Angular distance is measured in units of \(\Delta R \equiv \sqrt{(\Delta\eta)^{2} + (\Delta\phi)^{2}}\).}

The ATLAS experiment~\cite{PERF-2007-01, ATLAS-TDR-19, PIX-2018-001} at the LHC is a multipurpose particle detector
with a forward--backward symmetric cylindrical geometry and a near \(4\pi\) coverage in
solid angle.\footnote{\AtlasCoordFootnote}
It consists of an inner tracking detector surrounded by a thin superconducting solenoid
providing a \SI{2}{\tesla} axial magnetic field, electromagnetic and hadron calorimeters and a muon spectrometer.
A two-level trigger system is used to reduce the total event rate to
\SI{1}{\kHz} on average, depending on the data-taking conditions~\cite{TRIG-2016-01}.
An extensive software suite~\cite{ATL-SOFT-PUB-2021-001} is used in the reconstruction and analysis of real and simulated data,
in detector operations, and in the trigger and data acquisition systems of the experiment.
The events used in this analysis are selected using single-lepton triggers~\cite{TRIG-2018-01,TRIG-2018-05},
with either low thresholds for the lepton transverse momentum (\pt)\ and a lepton isolation requirement,
or higher thresholds, looser identification criteria and without any isolation requirement.
The lowest \pt\ threshold for muons is 20 (26)~\GeV, while for electrons the threshold is 24 (26)~\GeV\
for the data taken in 2015 (2016--2018).


\section{Event preselection}


Events are required to have at least one primary vertex, formed by two or more associated tracks with transverse momenta greater than $0.5$~\GeV.
The vertex with the highest sum of $\pt^2$ of associated tracks
is selected as the hard-scattering primary vertex.
Events with exactly one lepton (electrons or muons, denoted as $\ell$) or two oppositely charged leptons are considered in this analysis,
referred to as the \ljets\ channel and dilepton channel, respectively.
Electrons are identified using the `Tight' likelihood criterion~\cite{EGAM-2018-01}
and are required to have $\pt > 10$~\GeV\ and $|\eta| < 2.47$,
excluding those in the calorimeter barrel--endcap transition region ($1.37 < |\eta| < 1.52$).
Muons are selected with the `Medium' identification criterion~\cite{MUON-2018-03}
and are required to have $\pt > 10$~\GeV\ and $|\eta| < 2.5$.
Electrons (muons) are required to pass the `Gradient' (`Fixed-Cut-Tight-Track-Only') isolation requirements~\cite{EGAM-2018-01, MUON-2018-03}.
All leptons are required to originate from the primary vertex.
At least one of the leptons must have $\pt > 27$~\GeV\ and match the corresponding lepton used in the trigger decision.
In events with an $ee$ or $\mu\mu$ pair, the dilepton invariant mass is required to be above 15~\GeV\ and outside the $Z$ boson mass window of 83--99~\GeV.

This analysis targets events with high jet multiplicities, including $b$-quark jets
expected in the final state of \ttH\ and \tH\ events with a subsequent $H \rightarrow b\bar{b}$ decay.
Following the same procedure as Ref.~\cite{HIGG-2020-23}, jets are reconstructed from topological clusters of energy depositions in the calorimeter~\cite{PERF-2014-07,JETM-2018-05} using the anti-$k_t$ algorithm~\cite{Cacciari:2008gp, Fastjet} with a radius parameter of $R=0.4$.
The MV2c10 algorithm~\cite{FTAG-2018-01} is used to identify  (or `$b$-tag') jets containing $b$-hadrons.
By placing different selections on the MV2c10 discriminant, four working points are defined with average $b$-jet tagging efficiencies of 60\%, 70\%, 77\% and 85\% and different $c$- and light-jet rejection rates.
The corresponding efficiencies and rejection rates are calibrated to data~\cite{FTAG-2018-01,ATLAS-CONF-2018-001,ATLAS-CONF-2018-006}.
A pseudo-continuous b-tagging score is assigned to each jet.
A score of two, three, four and five is assigned if a jet passes the 85\%, 77\%, 70\% and 60\% working point, but fails the adjacent tighter one.
If a jet fails all working points, a score of one is assigned.
In the \ljets\ (dilepton) channel, events are required to have at least five (three) jets with $\pt > 25$~\GeV\ and $|\eta| < 2.5$,
and at least four (three) of the jets are required to be $b$-tagged at the 70\% efficiency working point.

The missing transverse momentum is reconstructed as the negative vector sum of the \pt\ of all selected objects in the event,
with an extra `soft term' built from additional tracks associated with the primary vertex~\cite{PERF-2016-07}.

The analysis also exploits the collimated decay topology from high-\pt\ Higgs bosons.
Jets with a radius parameter of $R=0.4$ are reclustered~\cite{Nachman:2014kla} using the anti-$k_t$ algorithm with a radius parameter of $R=1.0$.
The resulting jets are referred to as \textit{large}-$R$ jets.
The large-$R$ jets are required to have a mass larger than 50~\GeV, $\pt>200$~\GeV\
and at least two constituent jets with $R=0.4$.


\section{Signal and background modelling}

 
After applying the above selection criteria, background events are dominated by \ttbar production with additional jets (\ttjets),
that contain heavy-flavour hadrons ($b$- and $c$-hadrons).
Other processes contribute less than 10\% of the total expected background.
All background processes are estimated using Monte Carlo (MC) simulations,
closely following  Ref.~\cite{HIGG-2020-23}.

The simulated events were produced using the ATLAS detector simulation~\cite{SOFT-2010-01} based on \GEANT~\cite{Agostinelli:2002hh}.
To simulate the effects of multiple interactions in the same and neighbouring bunch crossings (pile-up),
additional interactions were generated using \PYTHIA[8.186]~\cite{Sjostrand:2014zea}
with a set of tuned parameters called the A3 tune~\cite{ATL-PHYS-PUB-2016-017}
and overlaid on the simulated hard-scatter event.
Simulated events are reweighted to match the pile-up conditions observed in the full Run 2 dataset.
All simulated event samples are processed through the same reconstruction algorithms and analysis chain as the data~\cite{ATL-SOFT-PUB-2021-001}.

Events in the simulated \ttjets\ background sample are categorised according to the flavour of the additional jets which do not originate from the top-quark decay.
The simulation of each set of backgrounds is treated independently as this allows for a more accurate modelling of \ttjets\ events.
The categorisation is based on `MC-truth jets' that are clustered with stable generated particles (with mean lifetime $\tau > 3 \times 10^{-11}$~s) in the final state
using the anti-$k_t$ algorithm with $R=0.4$.
MC-truth jets with $\pt > 15$~\GeV\ and $|\eta| < 2.5$ in the simulated events are used for the categorisation.
Their MC-truth flavour is determined by counting the number of $b$/$c$-hadrons contained within $\Delta R=0.4$ of the jet axis.
Events with at least one MC-truth jet containing $b$-hadrons not originating from a top-quark decay are labelled as \ttbin.
This can be further separated into subcomponents corresponding to \ttbexcl\ and \ttbbin.
Events failing to satisfy that criterion but with at least one MC-truth jet containing $c$-hadrons not originating from top-quark decay are labelled \ttcin.
The rest of the events are labelled as \ttlight.
The dominant \ttbin\ background is modelled using a sample of \ttbb\ events generated at next-to-leading order (NLO) in QCD in the four-flavour scheme,
with two additional massive $b$-quarks produced at the matrix element (ME) level.
The ME simulation was performed using the  \POWHEGBOXRES  ~generator and \OPENLOOPS ~\cite{Jezo:2018yaf,Cascioli:2011va,Denner:2016kdg,ttbbPowheg},
with the \textsc{NNPDF3.0nlo}~nf4~\cite{Ball:2014uwa} parton distribution function (PDF) set
and \PYTHIA[8.230]~\cite{Sjostrand:2014zea} with the A14 set of tuned parameters~\cite{ATL-PHYS-PUB-2014-021} for the simulation of the parton shower (PS) and hadronisation.
Given that the production rate of \ttbar\ with additional $b$-jets is
observed to be underestimated by the current predictions~\cite{TOPQ-2017-12,CMS-TOP-18-011},
the normalisation of the \ttbin\ background is determined from the analysed data without prior constraints.
The \ttcin\ and \ttlight\ backgrounds are modelled from a subset of an inclusive \ttjets\ sample generated at NLO in QCD
using \POWHEGBOX[v2]~\cite{Nason:2004rx,Frixione:2007vw,Alioli:2010xd,Hartanto:2015uka} as the ME generator
interfaced with \PYTHIA[8.230] for the PS and hadronisation.
This inclusive \ttjets\ sample is generated with the five-flavour scheme, where $c$- and $b$-quarks not originating from a top-quark decay are assumed to be massless.
Due to limited knowledge regarding  \ttcin\ production, an additional 100\% uncertainty is included in its normalisation.
Additionally, a prior uncertainty of 6\% is assigned to the inclusive \ttjets\ production cross-sections according to
the predicted inclusive \ttbar\ production cross-section at NNLO+NNLL~\cite{Beneke:2011mq,Cacciari:2011hy,Baernreuther:2012ws,Czakon:2012zr,Czakon:2012pz,Czakon:2013goa,Czakon:2011xx}.
Other background processes include the production of $W + $~jets, $Z + $~jets, $t\bar{t}W$, $t\bar{t}Z$, $tZq$, $tWZ$, $t\bar{t}t\bar{t}$ and $WW/WZ/ZZ$ events.
These are all subdominant and modelled from simulation as detailed in Ref.~\cite{HIGG-2020-23}.
A small fraction of events contain misidentified leptons or leptons originating from the decay of heavy-flavour hadrons.
The contribution from these events is found to be negligible in the \ljets\ channel.
In the dilepton channel, this small contribution is modelled using a simulation.

The signal processes, \ttH\ and \tH, are simulated with different values of $\alpha$ and \kprime.
All other parameters were fixed to their SM values, including the $H\to b\bar{b}$ branching ratio.
The alternative scenarios were simulated using the NLO Higgs Characterisation~\cite{Artoisenet:2013puc,deFlorian:2016spz} model
implemented in \MGNLO\ with FeynRules~\cite{Alloul:2013bka,Degrande:2011ua}.
With a few exceptions, all signal samples were generated using the \MGNLO[ 2.6.2 ]~\cite{Alwall:2014hca} generator at NLO in QCD
using the five-flavour scheme with the  \NNPDF[3.0nnlo] PDF set,
interfaced with \PYTHIA[8.230] with the A14 set of tune parameters for PS and hadronisation.
The SM \ttH\ events were simulated using \MGNLO[ 2.6.0 ].
The renormalisation and factorisation scales were set to $\sqrt[3]{m_{\text{T}}(t)\cdot m_{\text{T}}(\bar{t}) \cdot m_\text{T}(H)}$,
where $m_{\text{T}} = \sqrt{m^2 + p_\text{T}^2}$ is the transverse mass of a particle.
The cross-section is normalised to 507~fb from the fixed-order calculation including NLO QCD and electroweak corrections,
with an uncertainty of 3.6\% from variations in PDF and \alphas and 9.2\% due to variations of the renormalisation and factorisation scales~\cite{deFlorian:2016spz,PhysRevD.19.941,BEENAKKER2003151,PhysRevD.68.034022,ZHANG20141,frixione:2015}.
A $K$-factor of 1.1 is derived by taking the ratio of the cross-section from the above fixed-order calculation to that from \MGNLO,
and is applied to all \ttH\ samples with different values of $\alpha$ and \kprime.
For the \tH\ signal, two subprocesses, \tHjb\ and \tWH, are considered.
The \tHjb\ (\tWH) events were generated in the four(five)-flavour scheme using the  \NNPDF[3.0nnlo] ~nf4 (\NNPDF[3.0nnlo]) PDF set~\cite{Ball:2014uwa},
with the renormalisation and factorisation scales set to the generator's default.
The cross-sections for the \tHjb\ and \tWH\ samples are obtained directly from \MGNLO.
In the SM scenario, the cross-section for \tHjb\ and \tWH\ are 60.1~fb and 16.7~fb, respectively.
Variations of the renormalisation and factorisation scales, including the consideration of the flavour scheme choice for the \tHjb\ process, contribute 15\% and 6.7\% to the uncertainty of the cross-sections of \tHjb\ and \tWH respectively.
Similarly, variations of the PDFs and \alphas result in a 3.7\% and 6.3\% uncertainty in the \tHjb\ and \tWH\ cross-sections, respectively.
A diagram removal scheme~\cite{Frixione:2007zp} is applied in the simulation of the \tWH\ events in order to remove diagrams already included in the \ttH\ simulation.
 
The yields of \ttH\ and \tH\ signals are parameterised as a function of the model parameters by smoothly interpolating between generated MC samples with varying $\alpha$ and \kprime.
The parameterisation is performed in each analysis bin.
Two \ttH\ samples with alternative values of $\alpha$ were generated,
corresponding to a pure \(CP\)-odd interaction ($\alpha=90^\circ$) and maximal \(CP\)-odd/\(CP\)-even mixing ($\alpha=45^\circ$).
The \ttH\ yields, $N_{\ttH}(\kprime, \alpha)$, are parameterised using the SM sample and the pure \(CP\)-odd sample
as ${\kprime}^2 c_\alpha^2 N_{CP\textrm{-even}} + {\kprime}^2 s_\alpha^2 N_{CP\textrm{-odd}}$,
where $c_\alpha = \cos\alpha$, $s_\alpha = \sin\alpha$,
and $N_{CP\textrm{-even}}$ and $N_{CP\textrm{-odd}}$ are the expected yields predicted by the SM and the \(CP\)-odd \ttH\ simulations, respectively.
This was verified to be a good approximation using the maximal mixing sample ($\alpha=45^\circ$),
with the difference in any analysis bin smaller than the uncertainties due to the limited number of simulated events.
In the case of \tH, the interference between diagrams with \(CP\)-even and \(CP\)-odd $t-H$ and SM $W-H$ couplings are considered in the parameterisation,
assuming contributions from lowest order diagrams of \tHjb\ and \tWH\ processes.
The signal yield in each analysis bin is parameterised as $N_{tH}(\kprime, \alpha) = A{\kprime}^{2}c_\alpha^2+B{\kprime}^{2}s_\alpha^2+C\kprime c_\alpha+D\kprime s_\alpha +E{\kprime}^{2}c_\alpha s_\alpha + F$.
Coefficients $A$-$F$ are derived separately for each analysis bin,
by fitting to the yields predicted by multiple simulated samples with varying \kprime\ and $\alpha$.
The terms with $c_\alpha^2$ and $s_\alpha^2$ correspond to the contribution from $CP$-even and $CP$-odd $t$-$H$ coupling, respectively.
The terms at the first order of $c_\alpha$ and $s_\alpha$ account for potential interference effects between $CP$-even and $CP$-odd $t$-$H$ coupling and SM $W$-$H$ coupling contributions.
The term $F$ represents the contribution from only the SM $W$-$H$ coupling.
Ten samples generated with different values of $\alpha$ and \kprime\ in addition to the SM \tH\ sample are used for the parameterisation.
These samples include:
samples where $\kprime=1$ and $\alpha$ is set between 15\textdegree\ to 90\textdegree\ in steps of 15\textdegree,
samples with $\kprime =$ -1, 0.5, and 2 where $\alpha=0$\textdegree\ and an additional sample with
$\alpha=45$\textdegree\ and $\kprime=2$.
Uncertainties due to limited number of MC events in these simulated samples are considered when performing the parameterisation fit in each bin.
Good closure was observed: the largest $\chi^{2}$ per degree of freedom was 0.19 in any given bin.
Uncertainties pertaining to the parameterisation of either signal were found to have a negligible impact on the measured values of $\alpha$ and \kprime.


\section{Analysis strategy}

 
In order to optimise the analysis sensitivity, events satisfying the preselection criteria are categorised into orthogonal regions
in two steps.
In the first step, control regions (CR) and training regions (TR) are defined using requirements on jet multiplicity, $b$-tagging and large-$R$ jets.
The TRs are defined according to the expected numbers of objects from the decay of the signal events,
whilst the CRs with lower object multiplicities are signal depleted.
The TRs broadly contain the signals and are used to train various multivariate algorithms (MVA).
Dedicated observables are constructed in the TRs to enhance sensitivity to the top-Higgs Yukawa \(CP\) coupling.
In the second step, MVAs are used to divide the TRs into signal regions (SR) and additional CRs with relatively high and low signal purity, respectively.
Given the small contribution expected from \tH\ events, the categorisation, MVAs and \(CP\)-sensitive observables are optimised for the \ttH\ signal.
All regions labelled CR and SR are simultaneously fit to the data using either specific observables or simple yields as specified below.
Both steps are described in detail below.

The first step of categorisation adopts a strategy similar to that described in Ref.~\cite{HIGG-2020-23},
devised to separate the SM signal from the various backgrounds.
A `boosted' region, labelled as \TRboosted, is firstly defined in the \ljets\ channel by requiring the presence of a high-\pt\ Higgs boson candidate which is identified using a deep neural network (DNN).
The DNN is trained to identify the boosted Higgs boson candidates from among large-$R$ jets with $\pt > 300$~\GeV~\cite{HIGG-2020-23}.
A mixture of constituent jet masses, pseudo-continuous $b$-tagging scores and jet substructure observables~\cite{PERF-2012-02} are used as input features for the training.
Events failing this DNN selection defining the \TRboosted\ region are categorised into CRs and TRs according to the number of jets ($j$) and various $b$-tagging ($b$) requirements.
Events in the TRs are required to have at least the number of jets and $b$-tagged jets expected from the final state of the \ttH\ signal.
This results in four statistically independent regions in the dilepton channel, named \CR[hi]{3}{3}, \CR[lo]{4}{3}, \CR[hi]{4}{3} and \TR{4}{4},
and three regions in the \ljets\ channel, named \CR[lo]{5}{4}, \CR[hi]{5}{4} and \TR{6}{4}.
The yields of these regions enter the fit.
The requirements used to define all CRs and TRs are summarised in Table~\ref{tab:regions}.
Regions labelled with `hi' (`lo') have relatively higher (lower) fractions of events with true $b$-jets not from top-quark decays,
and are selected with tight (loose) $b$-tagging requirements.
The average $\Delta R$ separation between $b$-jets (\drbbav) is used as the observable which enters the fit for \CR[lo]{5}{4} and \CR[hi]{5}{4} regions as it better constrains the shape of the backgrounds.
All mentioned CRs have different fractions of \ttlight, \ttcin\ and \ttbin\ events and this helps to constrain the systematic uncertainties in each of these components.

\begin{table}[!htb]
\centering
\caption{
Definition of the CRs and TRs according to the number of jets and $b$-tagged jets using different $b$-tagging selection criteria, and the number of boosted Higgs boson candidates.
For CRs, the bottom row indicates the observables used in the fit to data in the corresponding regions.
For the  \TRboosted region, the $b$-tagged jets flagged with $^{\dagger}$ are not constituents of the boosted Higgs boson candidate.
Events must pass $N_{b\text{-tag}}$ requirements for each $b$-tagging selection criteria.
}
\small
\renewcommand{\arraystretch}{1.4}
\begin{tabular}{l@{\hspace{3pt}}lc@{\hspace{3pt}}c@{\hspace{3pt}}c@{\hspace{3pt}}c>{\hspace*{-5pt}}cc@{\hspace{3pt}}c@{\hspace{3pt}}c@{\hspace{3pt}}c}
\toprule
\hline
\multicolumn{2}{l}{\multirow{2}*{Region}} & \multicolumn{4}{c}{Dilepton}  && \multicolumn{4}{c}{\ljets} \\
& & \TR{4}{4} & \CR[hi]{4}{3} & \CR[lo]{4}{3} & \CR[hi]{3}{3} && \TR{6}{4} & \CR[hi]{5}{4} & \CR[lo]{5}{4} & \TRboosted \\
\midrule \midrule 
\multicolumn{2}{l}{$N_{\mathrm{jets}}$} &  \multicolumn{3}{c|}{$\geq4$} & $=3$ && $\geq6$  & \multicolumn{2}{|c|}{$=5$} & $\geq4$ \\
\cline{1-6} \cline{8-11}
\multirow{4}{*}{$N_{b\text{-tag}}$}        & @85\%  & \multicolumn{4}{c}{--} && \multicolumn{4}{c}{$\geq4$} \\
\cline{3-6} \cline{8-11}
& @77\% & \multicolumn{4}{c}{--} && \multicolumn{3}{c|}{--} & $\ge2^\dagger$ \\
\cline{3-6} \cline{8-11}
& @70\% & $\geq4$ & \multicolumn{3}{|c}{$=3$} && \multicolumn{3}{c|}{$\geq4$} & -- \\
\cline{3-6} \cline{8-11}
& @60\% & -- & $=3$ & $<3$ & $=3$ && -- & $\geq4$ & $<4$ & -- \\
\cline{1-6} \cline{8-11}
\multicolumn{2}{l}{$N_{\mathrm{boosted\ cand.}}$}  & \multicolumn{4}{c}{--} &&  \multicolumn{3}{c|}{0} & $\geq1$\\
\hline
\hline
\multicolumn{2}{l}{Fit observable} & -- & \multicolumn{3}{|c}{Yield} && -- & \multicolumn{2}{|c|}{\drbbav} & -- \\
\hline
\bottomrule
\end{tabular}
\label{tab:regions}
\end{table}
 
In the TRs, two sets of boosted decision trees (BDT) are trained: reconstruction BDTs and classification BDTs.
The former is trained to assign jets as coming from the decay of the Higgs boson or top quarks in \ttH\ events,
while the latter is trained to discriminate the \ttH\ signal against the backgrounds.
Both the reconstruction and classification BDTs are trained using simulated SM \ttH\ events.
It was tested that their performance is equally good for a pure \(CP\)-odd signal.
For both the reconstruction and classification BDTs, the training procedures are performed independently for each TR and are  identical to those used in Ref.~\cite{HIGG-2020-23}.
The reconstruction BDTs are trained to classify the correct combinations of jet assignments from random ones.
The training explores the relative positional information between pairs of objects,
and the invariant masses of object pairs and triplets that form $W$-boson and top-quark candidates.
In order to reconstruct the top-quark and Higgs boson candidates, for each event, all possible permutations of jet assignments are evaluated and the permutation with the highest BDT score is selected.
The reconstruction BDTs provide important information that improves the performance of the classification BDTs,
whilst allowing for the calculation of observables sensitive to the \(CP\) nature of the Yukawa coupling.
Classification BDT inputs include reconstruction BDT (DNN in the boosted channel) outputs, pseudo-continuous $b$-tagging discriminant scores of jets, and kinematic features,
such as angular separations and invariant masses of pairs of $b$-tagged jets.
The classification BDTs are used to further refine the TRs to define the final CRs and SRs, as detailed later.
The classification BDTs used in \TR{4}{4}, \TR{6}{4} and \TRboosted\ are henceforth denoted by \bdtdilep,
\bdtljets\ and \bdtboost, respectively.
 
Dedicated \(CP\)-sensitive observables are computed in \TR{4}{4} and \TR{6}{4} and are used in the fit
to determine the \(CP\) properties of the top-quark Yukawa coupling.
Two \(CP\) observables, $b_2$ and $b_4$~\cite{Gunion:1996xu,Ferroglia:2019qjy}, were found to provide the best discrimination in \TR{6}{4} of the \ljets\ channel and \TR{4}{4} of the dilepton channel, respectively.
They are defined as:
\begin{equation}
b_{2} = \frac{(\vec{p}_{1}\times \hat{z})\cdot(\vec{p}_{2}\times\hat{z})}{|\vec{p}_{1}||\vec{p}_{2}|},
\,\,\, \mathrm{ and } \,\,\,
b_{4} = \frac{(\vec{p}_{1}\cdot\hat{z})(\vec{p}_{2}\cdot\hat{z})}{|\vec{p}_{1}||\vec{p}_{2}|}, \nonumber
\end{equation}
where $\vec{p_i}$ with $i=1,2$ are the momentum three-vectors of the two top quarks in the events
and $\hat{z}$ is a unit vector in the direction of the beamline and defines the $z$-axis.
The $b_4$ observable exploits the enhanced production of top quarks travelling in opposite longitudinal directions
and closer to the beamline in \(CP\)-odd \ttH\ production.
The observable $b_2$ relies simultaneously on the smaller azimuthal separation of top quarks
and on their larger longitudinal fraction of momentum in \(CP\)-odd \ttH\ production.
The calculation of $b_2$ is performed in the \ttH\ rest frame~\cite{Ferroglia:2019qjy},  which enhances the discrimination power.
 
Computation of $b_2$ and $b_4$ requires the full reconstruction of both top quarks and the Higgs boson.
However, the reconstruction BDTs only resolve the hadronic part of the \ttH\ system.
In the \ljets\ channel, the missing transverse momentum is used as a proxy for the \pt\ of the undetected neutrino from the semileptonically decaying top quark.
The $z$ component of the neutrino four-momentum is obtained from a quadratic equation constructed from the lepton four-momentum and the missing transverse momentum, using as a constraint the leptonic $W$ boson's mass, assumed to be its on-shell value.
Both solutions of the quadratic equation are used to reconstruct the  top-quark mass,
and the one yielding a mass closer to 172.5~\GeV\ is chosen. In the case of a negative determinant,
a solution is obtained by setting the determinant to zero.
In the dilepton channel, the neutrino weighting technique is used to determine the four momenta of the two neutrinos~\cite{Abbott_1998,TOPQ-2016-04}.
Neutrino weighting provides a solution for reconstructing the \ttbar\ pair for 68\% of the events in \TR{4}{4}.
 
In contrast to the \TR{4}{4} and \TR{6}{4} regions, the \(CP\)-odd signals are strongly enhanced in comparison with the \(CP\)-even signals in the \TRboosted\ region.
The yields of \ttH\ with pure \(CP\)-even and \(CP\)-odd couplings are approximately equal in the \TRboosted\ region.
Additionally, the yield of the \tH\ signal with a pure \(CP\)-odd coupling is comparable to the \ttH\ signal yield.
The total \(CP\)-odd signal is therefore expected to be 50\% larger than a \(CP\)-even signal in this region.
Given the substantial sensitivity provided by the yield in this region,
the distribution of the classification BDT (\bdtboost) is used instead of a dedicated \(CP\)-sensitive observable.
 
In the second step of the categorisation, TRs are further refined to CRs and SRs according to the output of the reconstruction and classification BDTs.
A summary of the selections used to define the regions is detailed in Table~\ref{tab:sr_bdt}.
In \TRboosted, events below a classification BDT score of --0.05 are discarded to reduce contamination of \ttlight\ events.
\TR{4}{4} and \TR{6}{4} are further categorised, each into three regions, according to the classification BDT score.
The resulting regions have similar background compositions but different expected signal-to-background ratios ($S/B$).
The BDT thresholds are determined by optimising the sensitivity to the SM \ttH\ signal.
The three regions (one in \ljets\ and two in dilepton) with an $S/B>7$\% are referred to as SRs.
The remaining three regions (two in \ljets\ and one in dilepton) are used as additional CRs to constrain the modelling of the \(CP\) observables in the backgrounds events.
The highest $S/B$ in the resulting SRs is 22\% (10\%) for a pure \(CP\)-even (\(CP\)-odd) signal.
For \SR{4}{4} in the dilepton channel, $b_4$ cannot be calculated for events where the neutrino weighting fails to provide a solution.
These events are categorised as an additional region, \CR[\textrm{no-reco}]{4}{4},
where the difference in $\eta$ between the two leptons, $\Delta\eta_{\ell\ell}$, is used as a $CP$-sensitive observable instead~\cite{Demartin:2014fia}.

\begin{table}[htbp]
\centering
\caption{
Summary of the selections used to define SRs and CRs from the TRs, based on the classification BDT score.
In the boosted region, the selection requirement is applied and rejected events are removed entirely from further analysis.
In the dilepton channel, events with failed reconstruction due to absence of a real solution from the neutrino weighting are categorised into an additional region known as $\CR[\textrm{no-reco}]{4}{4}$.
The fitted discriminating variable in each region is indicated in the last column.
}
\begin{tabular}{l|c|c|c}
\toprule\hline
Channel (TR) & Final SRs and CRs    & Classification BDT selection & Fitted observable \\
\midrule
\multirow{4}{*}{Dilepton (\TR{4}{4})}
& \CR[\textrm{no-reco}]{4}{4} & --         & $\Delta\eta_{\ell\ell}$ \\
& \CR[ ]{4}{4} & $\bdtdilep\in [-1, -0.086)$     & $b_4$ \\
& \SR[1]{4}{4} & $\bdtdilep\in [-0.086, 0.186)$  & $b_4$ \\
& \SR[2]{4}{4} & $\bdtdilep\in [0.186, 1]$       & $b_4$ \\
\midrule
\multirow{3}{*}{\ljets\ (\TR{6}{4})}
& \CR[1]{6}{4} & $\bdtljets\in [-1, -0.128)$     & $b_2$ \\
& \CR[2]{6}{4} & $\bdtljets\in [-0.128, 0.249)$  & $b_2$ \\
& \SR[ ]{6}{4} & $\bdtljets\in [0.249, 1]$       & $b_2$ \\
\midrule
\ljets\ (\TRboosted) & \SRboosted    & $\bdtboost\in [-0.05, 1]$ & \bdtboost \\
\hline\bottomrule
\end{tabular}
\label{tab:sr_bdt}
\end{table}


\section{Systematic uncertainties}

Systematic uncertainties are assessed for three main sources:
theoretical modelling of the signal processes, background modelling which is dominated by the uncertainties in the \ttbin\ background
and experimental sources involving the (mis)identification rates and energy calibration of leptons, jets, $b$-jets and missing transverse momentum.
Uncertainties accounting for the limited number of events in all simulated samples are also considered.
Systematic variations can affect the overall yields, relative yields between analysis regions and shapes of observables.
 
Uncertainties associated with the modelling of the \ttH\ signals include variations due to initial and final state radiation (ISR and FSR),  choice of the NLO matching procedure as well as the PS and hadronisation model.
These uncertainties are evaluated using events generated with \POWHEGBOX\ + \PYTHIA[8], which are produced with the same PDF set and renormalisation and factorisation scales as the nominal \MGNLO\ + \PYTHIA[8] sample, unless otherwise specified.
Variations relative to the SM hypothesis are propagated to scenarios with alternative values of $\alpha$ and \kprime.
To estimate the uncertainty related to the amount of partonic ISR,
the renormalisation and factorisation scales in the ME and $\alpha_\text{S}^{\text{ISR}}$ in the PS are varied simultaneously~\cite{ATL-PHYS-PUB-2020-023}.
The impact of the FSR is evaluated by varying  $\alpha_\text{S}^{\text{FSR}}$ in the PS.
The impact of varying the PS and hadronisation models is estimated by comparing \ttH\ samples generated using \POWHEGBOX\ + \PYTHIA[ 8.230 ] with those generated from \POWHEGBOX\ + \HERWIG[ 7.04 ]~\cite{Bellm:2015jjp}.
The uncertainty due to the choice of NLO matching procedure is derived by directly comparing the \POWHEGBOX\ + \PYTHIA[8] sample with the nominal \MGNLO\ + \PYTHIA[8] sample.
The uncertainties in the modelling of \tH\ are estimated using the nominal sample generated using \MGNLO\ + \PYTHIA[8].
For each \tH\ subprocess (\tHjb\ and \tWH), two sources of modelling uncertainty are considered:
that associated with the description of PDFs, and the uncertainty due to missing higher-order QCD contributions.
The former is estimated from the standard deviation of the expected yields using 100 \NNPDF[3.0nlo] eigenvector PDF sets, in each analysis bin used to build the likelihood function.
The latter is estimated by coherently varying \muR ~and \muF  by factors of 0.5 and 2.

The most important uncertainties in the background estimation come from the modelling of the \ttbin\ background.
These uncertainties are designed to account for the choice of NLO matching procedure, PS and hadronisation model as well as the flavour scheme utilised in the \ttbin\ event generation.
An uncertainty in the ME-to-PS matching procedure is assessed
by comparing the \POWHEGBOX\ + \PYTHIA[8] sample with a sample generated using \MGNLO\ + \PYTHIA[8], both in the five-flavour scheme.
The variation by comparing these two samples is propagated to the nominal \ttbin\ sample generated with  \POWHEGBOXRES + \PYTHIA[8] in the four-flavour scheme.
This uncertainty is separated into three components that are treated independently:
one for the dilepton channel, another for the non-boosted regions in the \ljets\ channel, and a third for the \ljets\ boosted region.
This treatment is found to be important because it provides the fit with enough flexibility to cover the potential background mismodelling.
Uncertainties in the choice of the PS model are evaluated by comparing the nominal sample with the one produced with \POWHEGBOX\ + \HERWIG[7].
These uncertainties are treated in the same way as the uncertainty in the NLO matching procedure.
An additional source of systematic uncertainty is introduced to address the choice of flavour scheme used for the generation of the \ttbin\ events.
It is evaluated by comparing the nominal sample, generated in the four-flavour scheme using \POWHEG\ + \PYTHIA[8], with
that produced in the five-flavour scheme reweighted to remove differences in scale settings.
Uncertainties in ISR and FSR are estimated using the same procedure as used for the \ttH\ signals.
An uncertainty due to differences in relative fraction of \ttbexcl\ and \ttbbin\ subcomponents from different MC predictions is also considered.
Other uncertainties in \ttbin\ and uncertainties in other background components are treated identically to the procedure described in Ref.~\cite{HIGG-2020-23}.

Aside from the modelling uncertainties described above, experimental uncertainties are also considered.
These arise from the modelling of trigger, reconstruction, identification and isolation efficiencies,
as well as the calibration of energy and momentum scales for all physics objects, including electrons, muons, jets, $b$-tagged jets and \met.
Uncertainties in the measured integrated luminosity and in the modelling of additional $pp$ collisions are included.


\section{Results}

 
A binned profile likelihood fit is performed including all analysis regions simultaneously in order to determine the $\alpha$ and \kprime\ parameters.
The likelihood function, $\mathcal{L}(\alpha, \kprime, \pmb{\theta})$,
is constructed as the product of Poisson terms, with each term corresponding to an analysis bin.
The value of the likelihood varies according to the expected signal yields, as a function of $\alpha$ and \kprime, and background yields of the analysis bins,
as well as $\pmb{\theta}$, representing the nuisance parameters encoding the effects of the systematic uncertainties and a single parameter controlling the normalisation of the \ttbin\ background.
The nuisance parameters are constrained with Gaussian or log-normal functions.
The normalisation of the \ttbin\ background is controlled by an unconstrained parameter \kttb.
A profile likelihood ratio is used as the test statistic, following Ref.~\cite{Cowan:2010js}.
By scanning the value of the test statistic in grid points in  $\kappa'_t $ and $\alpha$, two-dimensional exclusion contours in the $(\kappa'_t , \alpha)$ plane are obtained.

Figure~\ref{fig:summary_regions} compares the observed yield of data in each analysis region with that expected after the fit to data (post-fit).
The expected yields for pure \(CP\)-even and \(CP\)-odd signals, normalised to the total data yields, are overlaid and shown with dashed lines in the top panels. These
illustrate the signal-to-background separation provided by the classification BDTs.
In the middle panel, the best-fit model is compared with the data by showing ratios of its value to the post-fit background prediction.
The post-fit model agrees well with the observed data.
In addition, the expected $S/B$ for pure \(CP\)-even and \(CP\)-odd signals are shown for both \ttH\ and \tH.
The post-fit yields for all backgrounds and the signals are summarised in Tables~\ref{tab:unblind_table_ljets}~and~\ref{tab:unblind_table_dilep}
for the \ljets\ and dilepton channels, respectively.
The expected yields of pure \(CP\)-even and \(CP\)-odd  signals are compared with the post-fit yields.
In all fitted regions, the best-fit signal yields are lower than their SM predictions.
The fitted value of \kttb\ is $1.30^{+0.09}_{-0.08}$, consistent with the value measured in Ref.~\cite{HIGG-2020-23}.
Figure~\ref{fig:SRs} shows the distributions of the fitted observables in the four SRs.
The post-fit predictions are in agreement with data.
Goodness-of-fit was evaluated using a likelihood ratio test,
comparing the likelihood value from the nominal fit with the one obtained from a saturated model
built with one free-floating normalisation factor for each fitted bin~\cite{Cousins2013GeneralizationOC}.
The probability that the post-fit prediction is compatible with the observed data is 80\%.
The pure \(CP\)-even and \(CP\)-odd signals are shown overlaid and normalised to the data yield to indicate the kinematic discrimination of the $b_2$ and $b_4$ observables.

\begin{figure}[!htb]
\centering
\includegraphics[width=0.95\textwidth]{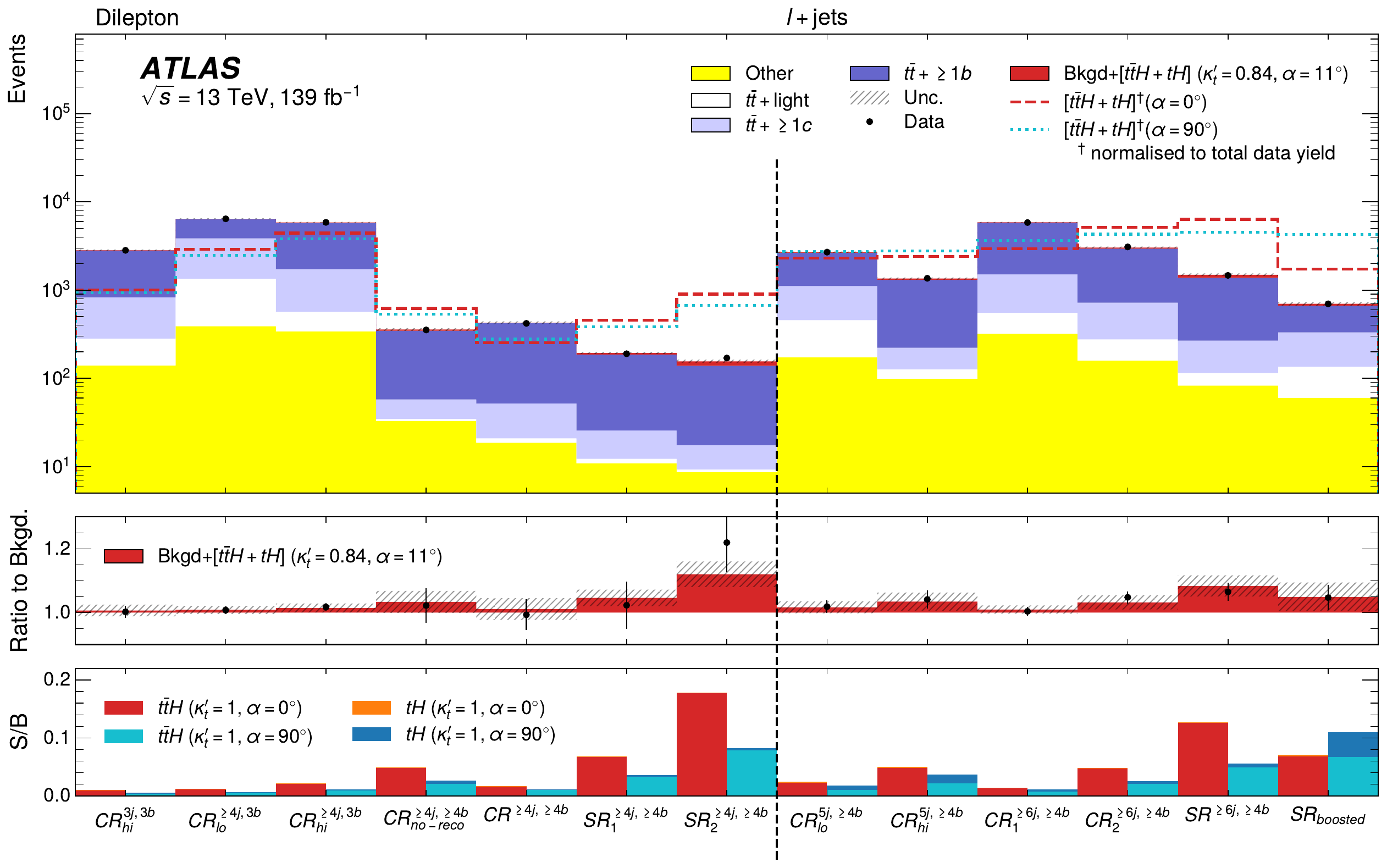}
\caption{
Yields calculated following a fit with \kprime\ and $\alpha$\ as free parameters, compared to the observed data in all analysis regions.
The different backgrounds and the signal are shown in coloured stack. The background component labelled ``other'' corresponds to the production of $W + $~jets, $Z + $~jets, $t\bar{t}W$, $t\bar{t}Z$, $tZq$, $tWZ$, $t\bar{t}t\bar{t}$ and $WW/WZ/ZZ$ events, as in Ref.~\cite{HIGG-2020-23}.
The dashed and dotted lines show the sum of \ttH + \tH\ signals for pure \(CP\)-even and \(CP\)-odd hypotheses normalised to the total data yields including all regions.
The hashed area around the prediction illustrates the total post-fit uncertainties.
In the middle panel, the best-fit model is compared with the data by showing ratios of its value to the post-fit background prediction.
The histogram represents the total post-fit model including the best-fit signals.
The hashed band represents the total post-fit uncertainty as a ratio to the background.
In the bottom panel, the $S/B$ is shown for pure \(CP\)-even and \(CP\)-odd signals, separately.
The histograms are shown as a stack of \ttH\ and \tH.
}
\label{fig:summary_regions}
\end{figure}

\begin{table}[!htb]
\centering
\sisetup{round-mode=places,round-precision=2,table-parse-only=true}
\caption{
The observed data yields and the expected signal and background yields in the \ljets\ channel.
The expected yields of pure $CP$-even and $CP$-odd \ttH\ and \tH\ signals, with $\kprime=1$, are shown at the top of the table.
The uncertainties in the pure $CP$-even and $CP$-odd \ttH\ and \tH\ signals are the total uncertainties before fitting to data.
Below that are shown the post-fit \ttH\ and \tH\ yields, corresponding to $\kprime=0.84$ and $\alpha=11^{\circ}$.
The following seven rows show the yields and uncertainties of individual background sources, where ``other'' corresponds to $W + $~jets, $Z + $~jets, $t\bar{t}W$, $t\bar{t}Z$, $tZq$, $tWZ$, $t\bar{t}t\bar{t}$ and $WW/WZ/ZZ$ events, as in Ref.~\cite{HIGG-2020-23}.
The row labelled `Total' represents the total signal plus background post-fit yields.
The uncertainties in the post-fit yields are evaluated from the post-fit nuisance parameters
as well as the post-fit uncertainties in the fitted free parameters ($\alpha$ and \kprime\ for the signals and \kttb\ for the \ttbin\ background)
that affect the corresponding processes.
The correlations amongst all fitted parameters are taken into account.
Due to these correlations the uncertainties on the total yields do not correspond to the quadrature sum  of uncertainties of individual signals and backgrounds.
\label{tab:unblind_table_ljets}}
\begin{tabular}{l|S|S|S|S|S|S}
\toprule
& {\CR[lo]{5}{4}} & {\CR[hi]{5}{4}} & {\CR[1]{6}{4}} & {\CR[2]{6}{4}} & {\SR[]{6}{4}} & {\SRboosted} \\
\midrule
\midrule
$\ttH~(1,0^{\circ})$       & 60\pm9     & 63\pm10    & 78\pm11    & 139\pm18   & 173\pm26   & 46\pm6    \\
$tH (1,0^{\circ})$         & 3.5\pm0.5  & 3.8\pm0.6  & 3.3\pm0.6  & 2.3\pm0.6  & 1.3\pm0.4  & 1.9\pm0.4 \\
\midrule
$\ttH~(1,90^{\circ})$      & 28\pm6     & 28\pm6     & 45\pm11    & 61\pm12    & 68\pm16    & 45\pm6    \\
$tH (1,90^{\circ})$        & 19.0\pm2.8 & 19.4\pm3.1 & 17.4\pm3.1 & 13.1\pm3.5 & 10\pm4     & 29\pm6    \\
\midrule
\midrule
$\ttH$ $(0.84,11^{\circ})$ & 40\pm30    & 41\pm31    & 50\pm40    & 90\pm70    & 110\pm80   & 30\pm22   \\
$tH$   $(0.84,11^{\circ})$ & 3\pm4      & 3.9\pm1.9  & 3.1\pm1.9  & 1.9\pm0.8  & 1.3\pm1.7  & 3\pm5     \\
\midrule
$t\bar{t} + \geq 1b$       & 1530\pm80  & 1090\pm60  & 4300\pm120 & 2220\pm120 & 1110\pm110 & 335\pm30  \\
$t\bar{t} + \geq 1c$       & 650\pm50   & 96\pm11    & 950\pm80   & 450\pm40   & 153\pm15   & 196\pm22  \\
$t\bar{t} + $ light        & 280\pm40   & 28\pm8     & 230\pm60   & 117\pm26   & 32\pm11    & 76\pm15   \\
Other                      & 173\pm30   & 99\pm20    & 320\pm50   & 159\pm21   & 83\pm11    & 60\pm11   \\
\midrule
Total                      & 2690\pm50  & 1350\pm40  & 5870\pm80  & 3040\pm70  & 1500\pm50  & 701\pm31  \\
\midrule
Data                       & 2696       & 1363       & 5837       & 3090       & 1470       & 699       \\
\bottomrule
\end{tabular}
\end{table}
 
\begin{table}[!htb]
\centering
\caption{
The observed data yields and the expected signal and background yields in the dilepton channel.
The expected yields of pure $CP$-even and $CP$-odd \ttH\ and \tH\ signals, with $\kprime=1$, are shown at the top of the table.
The uncertainties in the pure $CP$-even and $CP$-odd \ttH\ and \tH\ signals are the total uncertainties before fitting to data.
Below that are shown the post-fit \ttH\ and \tH\ yields, corresponding to $\kprime=0.84$ and $\alpha=11^{\circ}$.
The following seven rows show the yields and uncertainties of individual background sources, where ``other'' corresponds to $W + $~jets, $Z + $~jets, $t\bar{t}W$, $t\bar{t}Z$, $tZq$, $tWZ$, $t\bar{t}t\bar{t}$ and $WW/WZ/ZZ$ events, as in Ref.~\cite{HIGG-2020-23}.
The row labelled `Total' represents the total signal plus background post-fit yields.
The uncertainties in the post-fit yields are evaluated from the post-fit nuisance parameters
as well as the post-fit uncertainties in the fitted free parameters ($\alpha$ and \kprime\ for the signals and \kttb\ for the \ttbin\ background)
that affect the corresponding processes.
The correlations amongst all fitted parameters are taken into account.
Due to these correlations the uncertainties in the total yields do not correspond to the quadrature sum of uncertainties of individual signals and backgrounds.
\label{tab:unblind_table_dilep}}
\footnotesize
\sisetup{round-mode=places,round-precision=2,table-parse-only=true}
\begin{tabular}{l|S|S|S|S|S|S|S}
\toprule
& {\CR[hi]{3}{3}} & {\CR[lo]{4}{3}} & {\CR[hi]{4}{3}} & {\CR[\text{no-reco}]{4}{4}} & {\CR[]{4}{4}}  & {\SR[1]{4}{4}} & {\SR[2]{4}{4}} \\
\midrule
\midrule
$\ttH~(1,0^{\circ})$       & 26\pm4      & 79\pm8      & 120\pm12    & 16.9\pm2.1  & 6.9\pm1.1     & 12.5\pm1.5    & 24.8\pm2.9  \\
$tH (1,0^{\circ})$         & 1.12\pm0.13 & 0.90\pm0.13 & 1.74\pm0.20 & 0.19\pm0.08 & 0.087\pm0.035 & 0.100\pm0.033 & 0.09\pm0.06 \\
\midrule
$\ttH~(1,90^{\circ})$      & 10.6\pm1.6  & 35.6\pm3.5  & 54\pm5      & 7.2\pm0.9   & 4.3\pm0.6     & 6.1\pm0.7     & 10.9\pm1.3  \\
$tH (1,90^{\circ})$        & 5.4\pm0.6   & 7.0\pm1.0   & 10.7\pm1.2  & 1.8\pm0.8   & 0.48\pm0.19   & 0.48\pm0.16   & 0.5\pm0.4   \\
\midrule
\midrule
$\ttH$ $(0.84,11^{\circ})$ & 18\pm14     & 50\pm40     & 80\pm60     & 11\pm9      & 4.7\pm3.4     & 8\pm6         & 17\pm12     \\
$tH$   $(0.84,11^{\circ})$ & 0.9\pm0.5   & 1.0\pm1.9   & 1.5\pm1.3   & 0.17\pm0.16 & 0.068\pm0.016 & 0.08\pm0.14   & 0.07\pm0.09 \\
\midrule
$t\bar{t} + \geq 1b$       & 1990\pm80   & 2520\pm110  & 4040\pm130  & 288\pm15    & 371\pm16      & 160\pm8       & 122\pm11    \\
$t\bar{t} + \geq 1c$       & 550\pm50    & 2510\pm150  & 1160\pm90   & 23\pm4      & 31.1\pm2.5    & 13.4\pm1.6    & 8.2\pm1.0   \\
$t\bar{t} + $ light        & 143\pm27    & 960\pm130   & 230\pm40    & 1.7\pm0.4   & 2.3\pm0.8     & 1.4\pm0.8     & 0.57\pm0.25 \\
Other                      & 140\pm11    & 390\pm19    & 340\pm40    & 33\pm8      & 18.6\pm2.5    & 10.9\pm1.3    & 8.7\pm1.0   \\
\midrule
Total                      & 2840\pm50   & 6430\pm80   & 5850\pm80   & 358\pm12    & 428\pm15      & 194\pm5       & 156\pm6     \\
\midrule
Data                       & 2827        & 6429        & 5865        & 354         & 420           & 190           & 170         \\
\bottomrule
\end{tabular}
\end{table}

\begin{figure}[!htb]
\centering
\subfloat[]{
\includegraphics[width=0.4\textwidth]{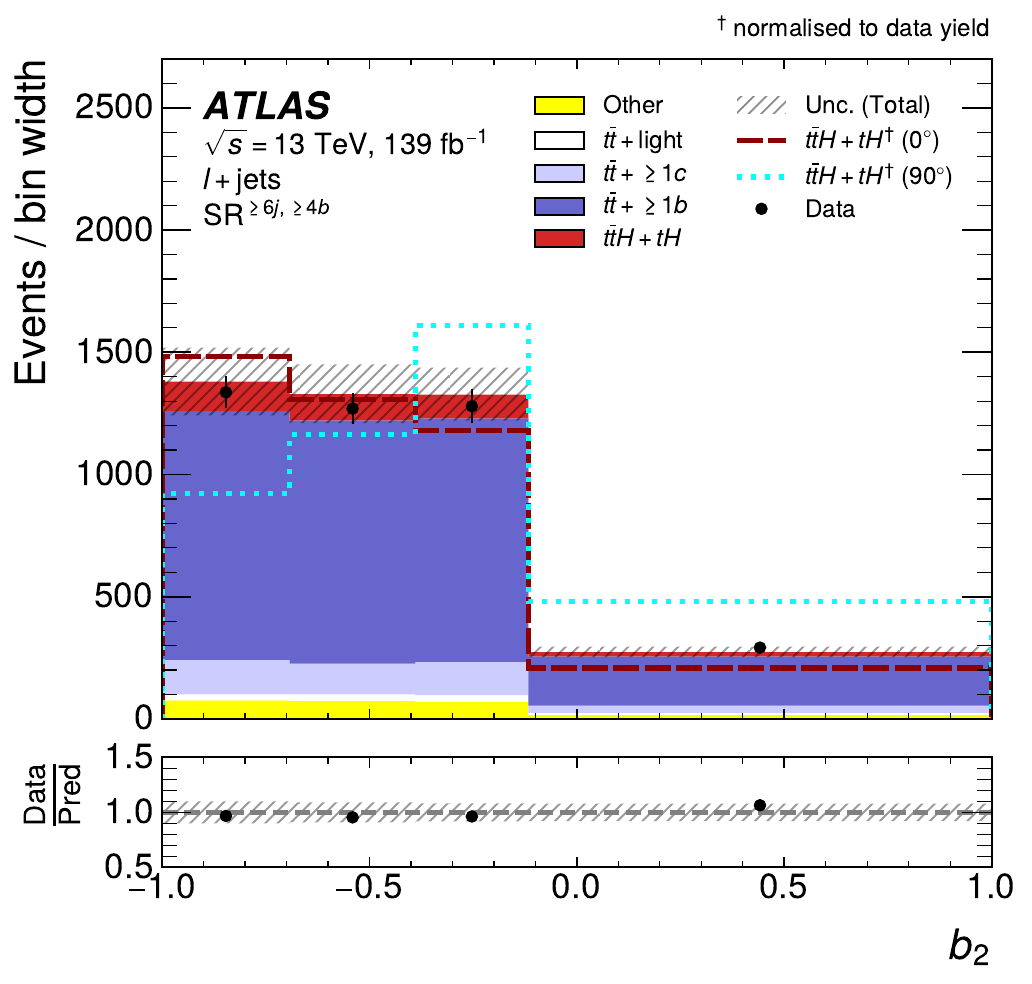}
}
\subfloat[]{
\includegraphics[width=0.4\textwidth]{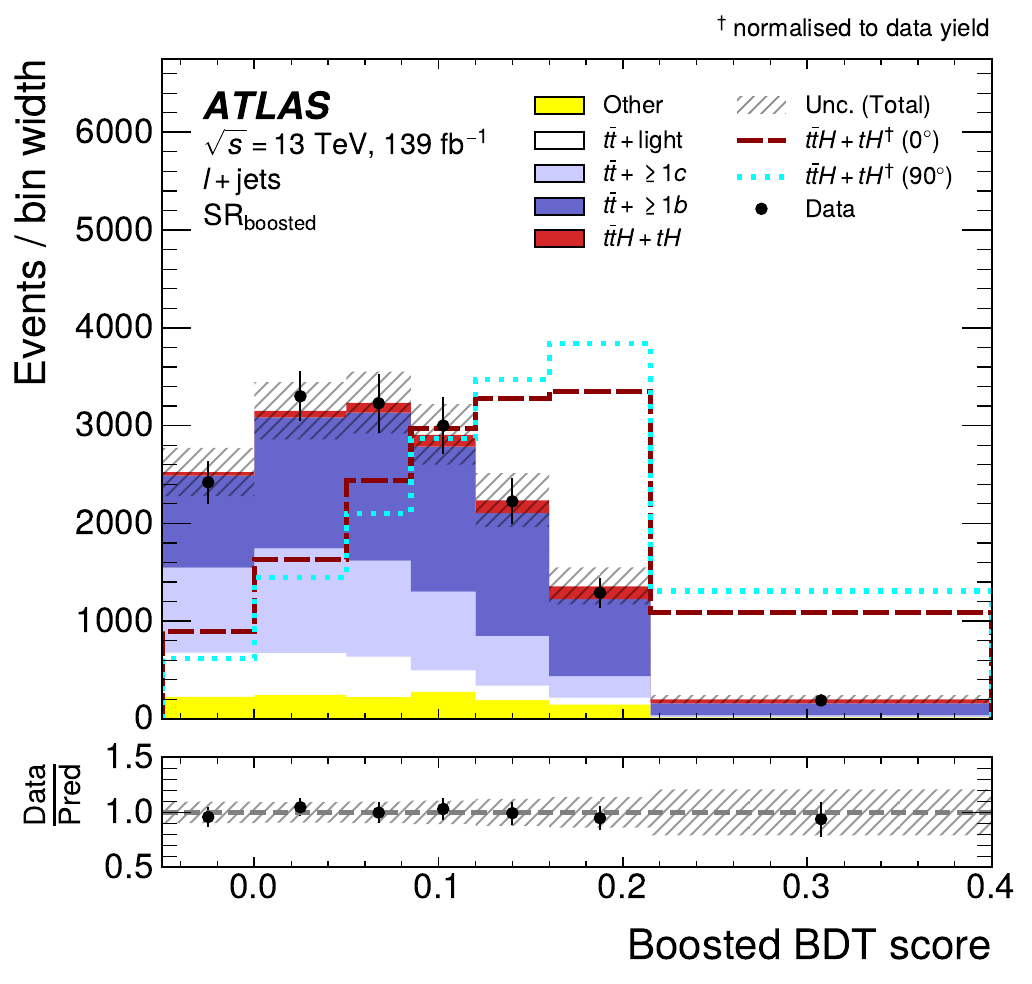}
}\\
\subfloat[]{
\includegraphics[width=0.4\textwidth]{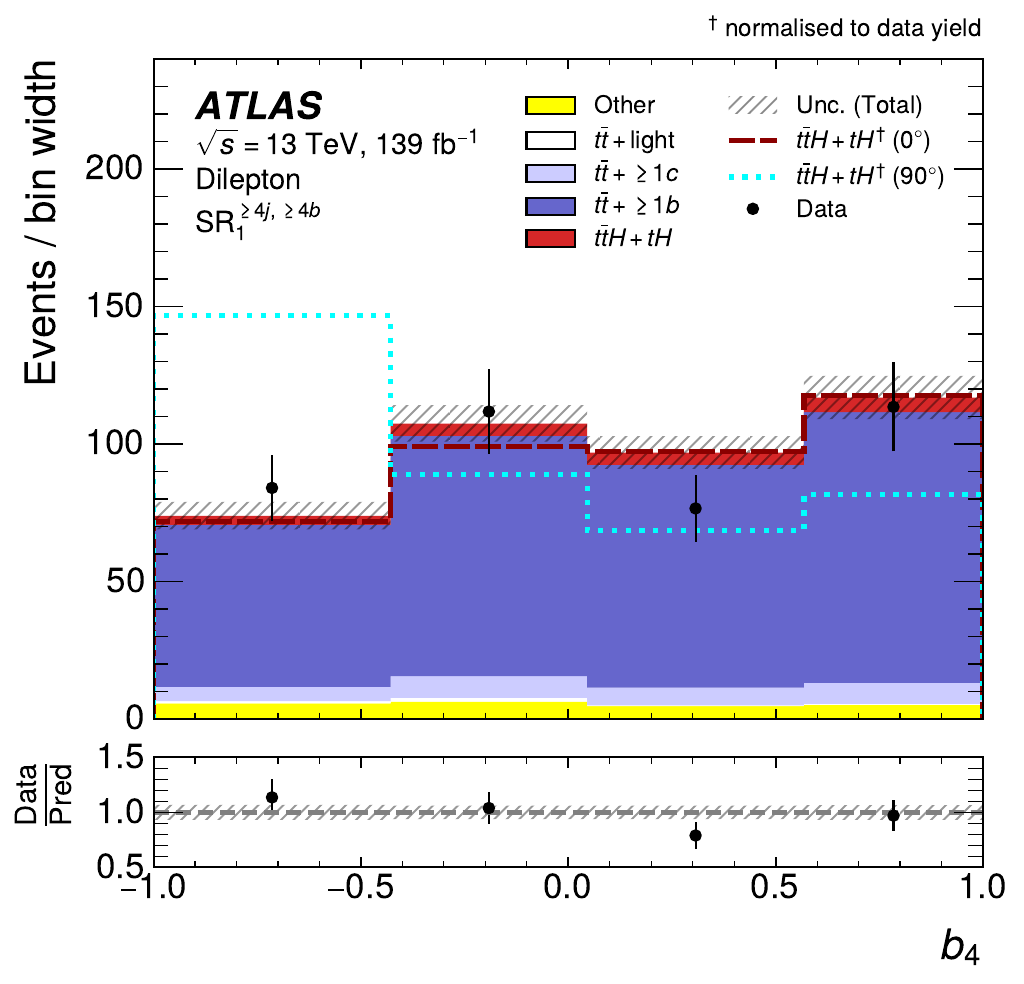}
}
\subfloat[]{
\includegraphics[width=0.4\textwidth]{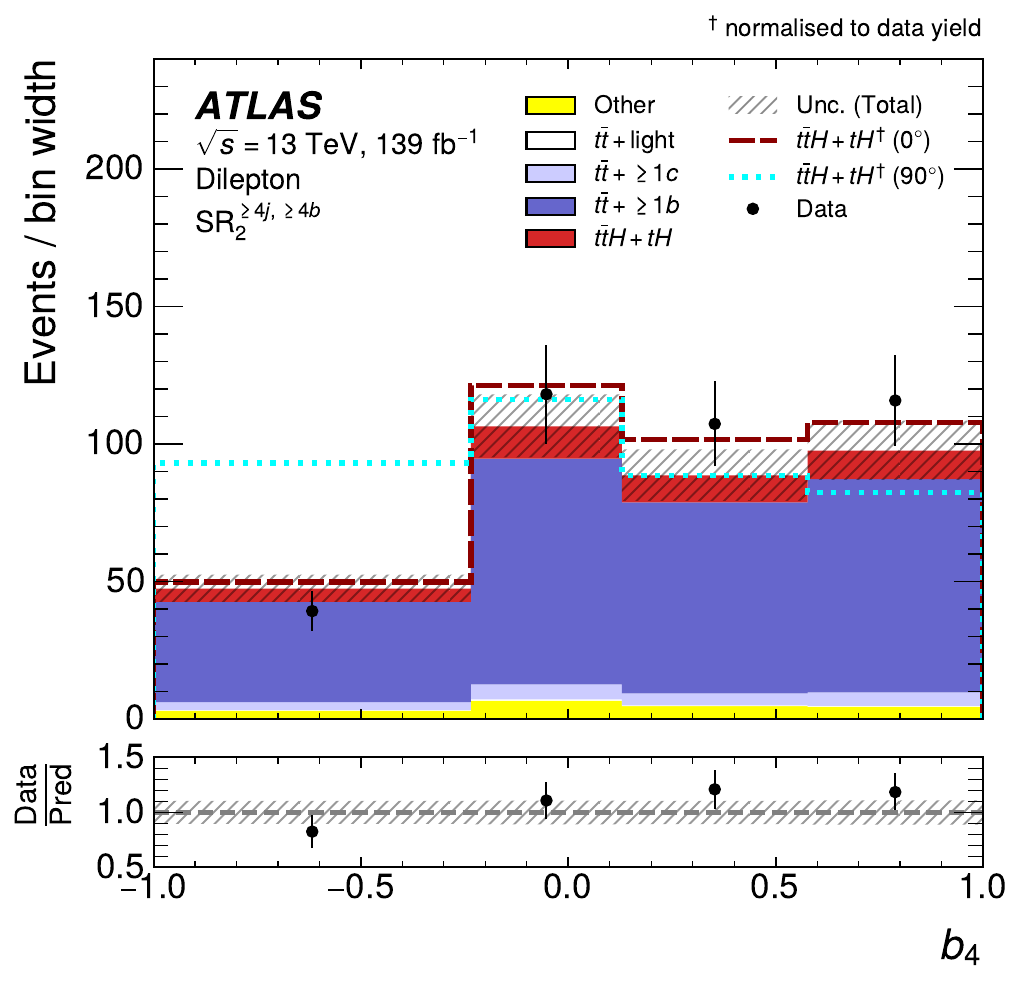}
}
\caption{
The distributions of the fitted variables in all signal regions.
The stacked histograms represent the predictions from a fit of signal and background to data with both \kprime\ and $\alpha$ as free parameters.
This is compared with data shown with black dots.
The solid red histogram shows the best-fit signal with $\alpha=11^\circ$ and $\kprime=0.84$.
The dashed and dotted lines show $\ttH+\tH$ signal predictions for pure \(CP\)-even and \(CP\)-odd hypotheses, respectively,
normalised to the total data yield per region in order to illustrate the shapes of the signal distribution.
The hashed area around the prediction illustrates the total post-fit uncertainties.
The lower panel shows the ratio of data to the predicted yields from a fit of signal and background in which \kprime\ and $\alpha$ are free parameters.
}
\label{fig:SRs}
\end{figure}

The best-fit values and the exclusion contours in $\alpha$ and \kprime\ are displayed in Figure~\ref{fig:LHscans} in the $(\kprime\cos\alpha , \kprime\sin\alpha)$ plane.
The best-fit value for the \(CP\) mixing angle $\alpha$ is $11^{\circ+52^{\circ}}_{~-73^{\circ}}$ and overall coupling strength $\kprime$ is $0.84^{+0.30}_{-0.46}$,
which  are in agreement with the SM expectations of $\alpha=0^{\circ}$ and $\kprime=1$.
The data disfavour the pure \(CP\)-odd hypothesis with a 1.2~$\sigma$ significance.
The significance of the observed \ttH\ and \tH\ signals over the background prediction is 1.3~$\sigma$.
The compatibility of this analysis with the \ttH\ cross-section measurement~\cite{HIGG-2020-23} was tested with the same parameter of interest:
a single free-floating signal strength, $\mu_{\ttH}$, controlling the normalisation of \ttH\ production in the SM scenario.
The \tH\ process was fixed to its SM prediction with an identical systematic model.
The compatibility is tested using the bootstrap technique~\cite{ATL-PHYS-PUB-2021-011}.
The difference in the measured $\mu_{\ttH}$ is sampled by fitting to toy datasets generated
by varying the event weights entering the Asimov dataset according to the Poisson fluctuations expected in data.
The measured values of $\mu_{\ttH}$ were found to be compatible within one standard deviation,
when accounting for the statistical correlations between the two measurements.

\begin{figure}[!htb]
\centering
\includegraphics[width=0.6\textwidth]{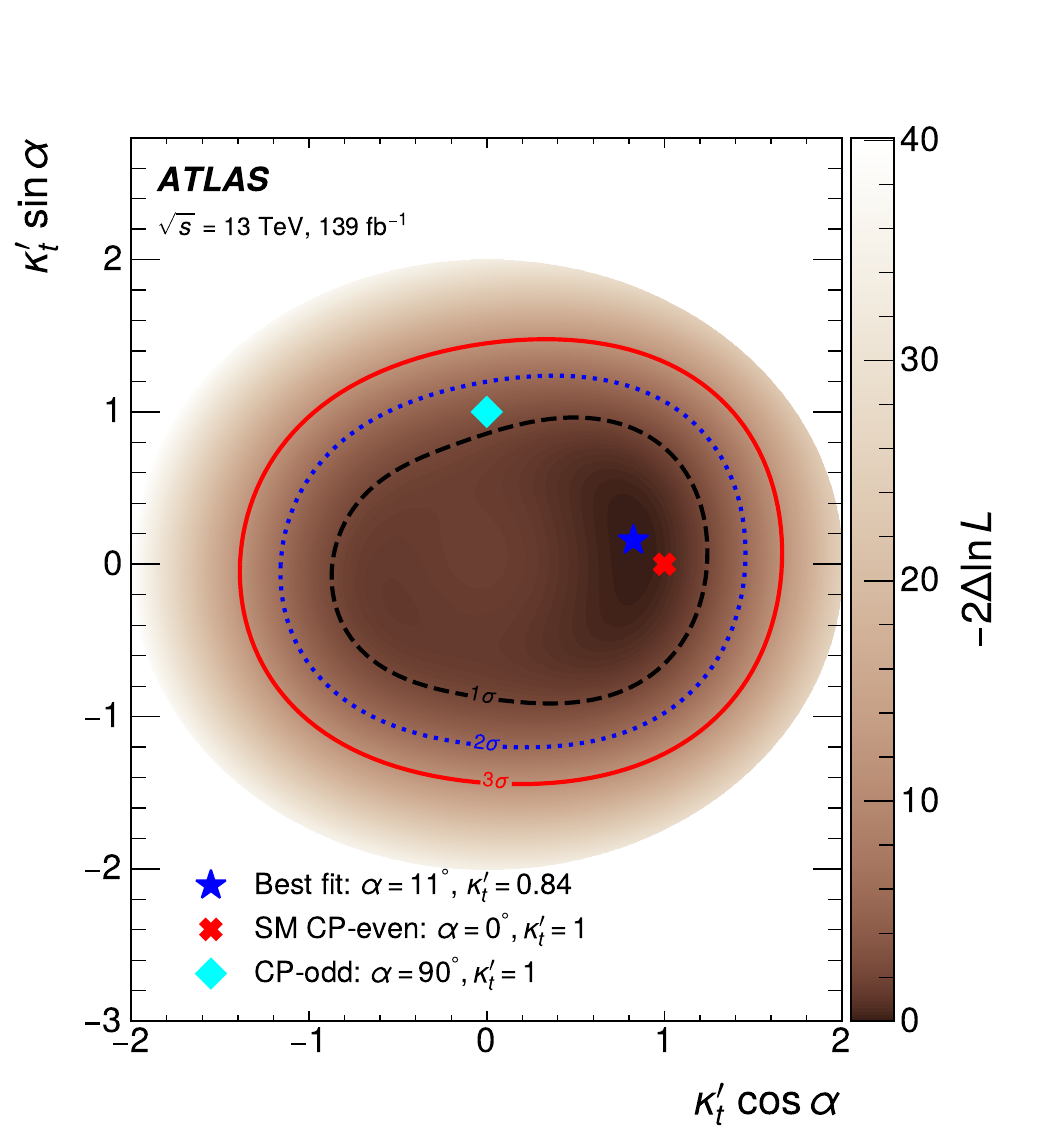}
\caption{
The observed exclusion contours in the $(\kprime\cos\alpha , \kprime\sin\alpha)$ plane.
Regions contained in the dashed, dotted and solid lines are compatible with the best-fit results at
1, 2 and 3~$\sigma$ standard deviations.
The cross (diamond) represents the \(CP\)-even (\(CP\)-odd) with \kprime=1 and the best-fit result is represented with a pentagram.
}
\label{fig:LHscans}
\end{figure}
 
The impact of a group of systematic uncertainties on $\alpha$ (\kprime) is assessed by fixing the nuisance parameters to their best fit values
and subtracting the subsequent $\alpha$ (\kprime) uncertainty in quadrature from the total $\alpha$ (\kprime) uncertainty.
The uncertainty in the measured value of $\alpha$ is dominated by \ttbin\ modelling uncertainties which contribute $^{+37^{\circ}}_{-51^{\circ}}$ to the overall uncertainty.
This is driven by: the NLO matching procedure between the ME and PS;
PS and hadronisation; and the choice of flavour scheme.
These uncertainties contribute $^{+22^{\circ}}_{-33^{\circ}}$, $^{+16^{\circ}}_{-24^{\circ}}$ and $^{+23^{\circ}}_{-37^{\circ}}$, respectively.
Smaller effects from the \ttbin\ modelling originate from the ISR uncertainty and the relative fractions of \ttbbin\ and \ttbexcl,
contributing $^{+14^{\circ}}_{-24^{\circ}}$ and $^{+14^{\circ}}_{-21^{\circ}}$.
The \ttcin\ modelling uncertainties contribute only $^{+6.6^{\circ}}_{-11^{\circ}}$ to the uncertainty in $\alpha$.
The 100\% \ttcin\ normalisation uncertainty is constrained to 50\% with a pull of 0.6~$\sigma$, and has negligible impact on the fitted $\alpha$ and \kprime.
Through a correlation with $\alpha$, the measured \kprime\ contributes
$^{+17^{\circ}}_{-33^{\circ}}$ to the $\alpha$ uncertainty.
Experimental uncertainties are smaller than the \ttbin\ modelling uncertainties.
The statistical uncertainty is $ ^{+32^{\circ}}_{-49^{\circ}}$.

\FloatBarrier
 
\section{Summary}
 
In conclusion, the \(CP\) properties of the top-quark's Yukawa coupling to the Higgs boson are probed in  \ttH\ and \tH\ production with $H\rightarrow b\bar{b}$ decays, which had not been studied before. Dedicated \(CP\)-sensitive variables  relying on angular separations between reconstructed top quarks or lepton candidates were used directly.
Assuming the SM branching ratio for the Higgs boson decay, the best-fit values of the \(CP\)-mixing angle and the overall coupling strength are
$\alpha = 11^{\circ+52^{\circ}}_{~-73^{\circ}}$ and $\kprime = 0.84^{+0.30}_{-0.46}$.
These values can be compared with the expected allowed 1~$\sigma$ ranges of $\alpha$ and \kprime,
obtained using Asimov datasets constructed with either a pure \(CP\)-even or -odd signal.
For a \(CP\)-even scenario $\alpha\in[-180^{\circ}, -173^{\circ}]\cup[-50^{\circ}, 52^{\circ}]\cup[171^{\circ}, 180^{\circ}]$ and $\kprime = 1.00^{+0.29}_{-0.27}$,
whilst for a pure \(CP\)-odd scenario $\alpha\in[-157^{\circ}, -41^{\circ}]\cup[43^{\circ}, 157^{\circ}]$ and $\kprime = 1.00^{+0.22}_{-0.33}$.
The sensitivity of this measurement is driven by the systematic uncertainties.
 
These results complement previous ATLAS measurements in the $H\rightarrow \gamma \gamma$ decay channel and will allow for a future combined measurement of the \(CP\) properties of the top-quark Yukawa coupling.
Due to the tree-level sensitivity and the high $H\rightarrow b\bar{b}$ branching ratio, it can be expected that future measurements in the \ttH\ and \tH\ channels will become quite sensitive to the \(CP\) properties of the top-quark Yukawa coupling. Additional LHC data and a better theoretical understanding of the \ttbin\ process will be essential ingredients in order to achieve this sensitivity.


\section*{Acknowledgements}


We thank CERN for the very successful operation of the LHC, as well as the
support staff from our institutions without whom ATLAS could not be
operated efficiently.
 
We acknowledge the support of
ANPCyT, Argentina;
YerPhI, Armenia;
ARC, Australia;
BMWFW and FWF, Austria;
ANAS, Azerbaijan;
CNPq and FAPESP, Brazil;
NSERC, NRC and CFI, Canada;
CERN;
ANID, Chile;
CAS, MOST and NSFC, China;
Minciencias, Colombia;
MEYS CR, Czech Republic;
DNRF and DNSRC, Denmark;
IN2P3-CNRS and CEA-DRF/IRFU, France;
SRNSFG, Georgia;
BMBF, HGF and MPG, Germany;
GSRI, Greece;
RGC and Hong Kong SAR, China;
ISF and Benoziyo Center, Israel;
INFN, Italy;
MEXT and JSPS, Japan;
CNRST, Morocco;
NWO, Netherlands;
RCN, Norway;
MEiN, Poland;
FCT, Portugal;
MNE/IFA, Romania;
MESTD, Serbia;
MSSR, Slovakia;
ARRS and MIZ\v{S}, Slovenia;
DSI/NRF, South Africa;
MICINN, Spain;
SRC and Wallenberg Foundation, Sweden;
SERI, SNSF and Cantons of Bern and Geneva, Switzerland;
MOST, Taiwan;
TENMAK, T\"urkiye;
STFC, United Kingdom;
DOE and NSF, United States of America.
In addition, individual groups and members have received support from
BCKDF, CANARIE, Compute Canada and CRC, Canada;
PRIMUS 21/SCI/017 and UNCE SCI/013, Czech Republic;
COST, ERC, ERDF, Horizon 2020, ICSC-NextGenerationEU and Marie Sk{\l}odowska-Curie Actions, European Union;
Investissements d'Avenir Labex, Investissements d'Avenir Idex and ANR, France;
DFG and AvH Foundation, Germany;
Herakleitos, Thales and Aristeia programmes co-financed by EU-ESF and the Greek NSRF, Greece;
BSF-NSF and MINERVA, Israel;
Norwegian Financial Mechanism 2014-2021, Norway;
NCN and NAWA, Poland;
La Caixa Banking Foundation, CERCA Programme Generalitat de Catalunya and PROMETEO and GenT Programmes Generalitat Valenciana, Spain;
G\"{o}ran Gustafssons Stiftelse, Sweden;
The Royal Society and Leverhulme Trust, United Kingdom.
 
The crucial computing support from all WLCG partners is acknowledged gratefully, in particular from CERN, the ATLAS Tier-1 facilities at TRIUMF (Canada), NDGF (Denmark, Norway, Sweden), CC-IN2P3 (France), KIT/GridKA (Germany), INFN-CNAF (Italy), NL-T1 (Netherlands), PIC (Spain), ASGC (Taiwan), RAL (UK) and BNL (USA), the Tier-2 facilities worldwide and large non-WLCG resource providers. Major contributors of computing resources are listed in Ref.~\cite{ATL-SOFT-PUB-2023-001}.


\printbibliography

\clearpage
 
 
\begin{flushleft}
\hypersetup{urlcolor=black}
{\Large The ATLAS Collaboration}

\bigskip

\AtlasOrcid[0000-0002-6665-4934]{G.~Aad}$^\textrm{\scriptsize 101}$,
\AtlasOrcid[0000-0002-5888-2734]{B.~Abbott}$^\textrm{\scriptsize 119}$,
\AtlasOrcid[0000-0002-7248-3203]{D.C.~Abbott}$^\textrm{\scriptsize 102}$,
\AtlasOrcid[0000-0002-1002-1652]{K.~Abeling}$^\textrm{\scriptsize 55}$,
\AtlasOrcid[0000-0002-8496-9294]{S.H.~Abidi}$^\textrm{\scriptsize 29}$,
\AtlasOrcid[0000-0002-9987-2292]{A.~Aboulhorma}$^\textrm{\scriptsize 35e}$,
\AtlasOrcid[0000-0001-5329-6640]{H.~Abramowicz}$^\textrm{\scriptsize 150}$,
\AtlasOrcid[0000-0002-1599-2896]{H.~Abreu}$^\textrm{\scriptsize 149}$,
\AtlasOrcid[0000-0003-0403-3697]{Y.~Abulaiti}$^\textrm{\scriptsize 116}$,
\AtlasOrcid[0000-0003-0762-7204]{A.C.~Abusleme~Hoffman}$^\textrm{\scriptsize 136a}$
% ...
\bigskip
\\

% ...
$^{29}$Physics Department, Brookhaven National Laboratory, Upton NY; United States of America.\\
% ...
$^{35}$$^{(a)}$Facult\'e des Sciences Ain Chock, R\'eseau Universitaire de Physique des Hautes Energies - Universit\'e Hassan II, Casablanca;$^{(b)}$Facult\'{e} des Sciences, Universit\'{e} Ibn-Tofail, K\'{e}nitra;$^{(c)}$Facult\'e des Sciences Semlalia, Universit\'e Cadi Ayyad, LPHEA-Marrakech;$^{(d)}$LPMR, Facult\'e des Sciences, Universit\'e Mohamed Premier, Oujda;$^{(e)}$Facult\'e des sciences, Universit\'e Mohammed V, Rabat;$^{(f)}$Institute of Applied Physics, Mohammed VI Polytechnic University, Ben Guerir; Morocco.\\
% ...
$^{55}$II. Physikalisches Institut, Georg-August-Universit\"{a}t G\"ottingen, G\"ottingen; Germany.\\
% ...
$^{101}$CPPM, Aix-Marseille Universit\'e, CNRS/IN2P3, Marseille; France.\\
$^{102}$Department of Physics, University of Massachusetts, Amherst MA; United States of America.\\
% ...
$^{116}$Department of Physics, New York University, New York NY; United States of America.\\
% ...
$^{119}$Homer L. Dodge Department of Physics and Astronomy, University of Oklahoma, Norman OK; United States of America.\\
% ...
$^{136}$$^{(a)}$Departamento de F\'isica, Pontificia Universidad Cat\'olica de Chile, Santiago;$^{(b)}$Millennium Institute for Subatomic physics at high energy frontier (SAPHIR), Santiago;$^{(c)}$Instituto de Investigaci\'on Multidisciplinario en Ciencia y Tecnolog\'ia, y Departamento de F\'isica, Universidad de La Serena;$^{(d)}$Universidad Andres Bello, Department of Physics, Santiago;$^{(e)}$Instituto de Alta Investigaci\'on, Universidad de Tarapac\'a, Arica;$^{(f)}$Departamento de F\'isica, Universidad T\'ecnica Federico Santa Mar\'ia, Valpara\'iso; Chile.\\
% ...
$^{149}$Department of Physics, Technion, Israel Institute of Technology, Haifa; Israel.\\
$^{150}$Raymond and Beverly Sackler School of Physics and Astronomy, Tel Aviv University, Tel Aviv; Israel.\\
% ...
$^{a}$ Also Affiliated with an institute covered by a cooperation agreement with CERN.\\
$^{b}$ Also at An-Najah National University, Nablus; Palestine.\\
$^{c}$ Also at Borough of Manhattan Community College, City University of New York, New York NY; United States of America.\\
$^{d}$ Also at Bruno Kessler Foundation, Trento; Italy.\\
$^{e}$ Also at Center for High Energy Physics, Peking University; China.\\
$^{f}$ Also at Centro Studi e Ricerche Enrico Fermi; Italy.\\
% ...

\end{flushleft}


\end{document}